\pdfoutput=1

\documentclass[journal]{vgtc}         % review (journal style)
\ifpdf%                                % if we use pdflatex
  \pdfoutput=1\relax                   % create PDFs from pdfLaTeX
  \usepackage{graphicx}                % allow us to embed graphics files
  \DeclareGraphicsExtensions{.pdf,.png,.jpg,.jpeg} % for pdflatex we expect .pdf, .png, or .jpg files
\else%                                 % else we use pure latex
  \usepackage{graphicx}                % allow us to embed graphics files
  \DeclareGraphicsExtensions{.eps}     % for pure latex we expect eps files
\fi%

\graphicspath{{figures/}{pictures/}{images/}{./}} % where to search for the images

\usepackage{microtype}                 % use micro-typography (slightly more compact, better to read)
\PassOptionsToPackage{warn}{textcomp}  % to address font issues with \textrightarrow
\usepackage{textcomp}                  % use better special symbols
\usepackage{mathptmx}                  % use matching math font
\usepackage{times}                     % we use Times as the main font
\usepackage{cite}                      % needed to automatically sort the references
\usepackage{tabu}                      % only used for the table example
\usepackage{booktabs}                  % only used for the table example
\usepackage{color,soul}
\usepackage{fixltx2e}
\usepackage{tabularx}
\usepackage{multirow}

\usepackage{makecell}

\usepackage{xspace}
\newcommand{\numpapersgleicher}{110\xspace}
\newcommand{\numtotalpapers}{401\xspace}
\newcommand{\numvenues}{50\xspace}
\newcommand{\numpapers}{127\xspace}
\newcommand{\numdesigns}{245\xspace}
\newcommand{\avedesigns}{1.9\xspace}
\newcommand{\numgeneralpapers}{112\xspace}
\newcommand{\numstudypapers}{15\xspace}
\newcommand{\numstudytasks}{40\xspace}
\newcommand{\numguidelines}{seven\xspace}
\newcommand{\change}[1]{\textcolor{black}{#1}}
\definecolor{codeblue}{rgb}{0.306, 0.475, 0.655}
\newcommand{\code}[1]{\textcolor{codeblue}{#1}}
\onlineid{1372}

\vgtccategory{Research}
\vgtcpapertype{theory/model}

\title{Comparative Layouts Revisited:\protect\\Design Space, Guidelines, and Future Directions}

\author{Sehi L'Yi, Jaemin Jo, and Jinwook Seo}
\authorfooter{
\item
 Sehi L'Yi is with Harvard Medical School. E-mail: sehi\_lyi@hms.harvard.edu.
\item
 Jaemin Jo is with Sungkyunkwan University. E-mail: jmjo@skku.edu.
\item
 Jinwook Seo is with Seoul National University. E-mail: jseo@snu.ac.kr.
}

\shortauthortitle{L'Yi \MakeLowercase{\textit{et al.}}: Comparative Layouts Revisited}
\abstract{
We present a systematic review on three comparative layouts---\textit{juxtaposition}, \textit{superposition}, and \textit{explicit-encoding}---which are information visualization (InfoVis) layouts designed to support comparison tasks.
For the last decade, these layouts have served as fundamental idioms in designing many visualization systems.
However, we found that the layouts have been used with inconsistent terms and confusion, and the lessons from previous studies are fragmented.
The goal of our research is to distill the results from previous studies into a consistent and reusable framework.
We review \numpapers research papers, including \numstudypapers papers with quantitative user studies, which employed comparative layouts.
We first alleviate the ambiguous boundaries in the design space of comparative layouts by suggesting lucid terminology (e.g., \textit{chart-wise} and \textit{item-wise} juxtaposition).
We then identify the diverse aspects of comparative layouts, such as the advantages and concerns of using each layout in the real-world scenarios and researchers' approaches to overcome the concerns.
Building our knowledge on top of the initial insights gained from the Gleicher et al.'s survey \cite{gleicher2011visualcomparison}, we elaborate on relevant empirical evidence that we distilled from our survey (e.g., the actual effectiveness of the layouts in different study settings) and identify novel facets that the original work did not cover (e.g., the familiarity of the layouts to people).
Finally, we show the consistent and contradictory results on the performance of comparative layouts and offer practical implications for using the layouts by suggesting trade-offs and \numguidelines actionable guidelines.% of using the layouts.
}

\CCScatlist{ % not used in journal version
 \CCScat{K.6.1}{Management of Computing and Information Systems}%
{Project and People Management}{Life Cycle};
 \CCScat{K.7.m}{The Computing Profession}{Miscellaneous}{Ethics}
}

\vgtcinsertpkg

\begin{document}


\firstsection{Introduction}

\maketitle

A decade ago, Gleicher et al. \cite{gleicher2011visualcomparison} suggested three primitive information visualization (InfoVis) layouts that support comparison tasks---\textit{juxtaposition}, \textit{superposition}, and \textit{explicit-encoding}---based on their literature survey on more than 100 research papers
(\autoref{figure:primitive_layouts}).
These layouts have served as fundamental idioms for designing comparative visualizations in diverse areas such as radiology \cite{song2016gazedx}, biology \cite{van2014bloodflow}, and geology \cite{alcalde2017framing}.
The layouts have also been popular in academia, as the number of papers citing the comparative layouts shows rapid growth.

To develop a better understanding of comparative layouts, researchers have attempted to study the effectiveness of the three layouts by conducting user studies and adopting the layouts to specific domains.
Gleicher et al. \cite{gleicher2011visualcomparison} initially discussed the potential strength and weakness of comparative layouts in terms of scalability, cognitive cost, and task performance, followed by many user studies in the Human--Computer Interaction (HCI) field \cite{tominski2012natural,jardine2019perceptual,liu2015effects,lobo2015interactivemap,naragino2017weightedfreetrees,ondov2018face,sambasivan2013visualizing,schmidt2013vaico,srinivasan2018whatsthedifference}.
Ondov et al. \cite{ondov2018face}, for example, compared the variants of juxtaposition and superposition, such as using adjacent, symmetric, and animated arrangements, in identifying the maximum change and correlation between two visualizations.

However, we find the lessons and practical findings from those previous studies fragmented, sometimes even with inconsistent terms.
For example, we encounter several visualizations techniques (e.g., variants of bar charts or heatmaps) that are inconsistently regarded as either juxtaposition or superposition \cite{alper2013weighted,srinivasan2018whatsthedifference,ondov2018face,zhao2015matrixwave,sambasivan2013visualizing,kim20173d4ddata}.
Moreover, contrary to the general consensus that superposition is more effective for small differences \cite{cruz2018interactive,heimerl2018multiclassscatter,caruso2017creating}, recent studies show that juxtaposition can be more effective for some tasks, such as comparing global characteristics between two bar charts \cite{ondov2018face,jardine2019perceptual}.

We present a systematic review on three comparative layouts with \numpapers research papers, including {\numstudypapers} papers with quantitative user studies, which employed the layouts.
\change{There are several studies that explored the design space of visualization arrangement.
Javed et al.~\cite{javed2012exploring} classified four design choices for composite visualizations, and Chen et al.~\cite{chen2014visual} suggested a conceptual framework for overlaying visualizations.
The main difference of our work compared with the prior studies is that we base our work on researchers' empirical findings from a wider-range of research papers.}
We combine and systematize the insights researchers gained \textit{in the wild}, for example, during a visualization design process in collaboration with data analysts or in evaluation with the actual users, and distill these studies into a consistent and reusable framework.

We first alleviate the unambiguous boundaries between comparative layouts using lucid classification to give implications in a more systematic and precise manner (e.g., \textit{chart-wise} and \textit{item-wise juxtaposition}; \autoref{figure:layouts}, \autoref{figure:num_layouts}, and \autoref{figure:layout-count}).
We then identify the diverse aspects of each layout (\autoref{table:justifications}), such as the advantages and concerns of using them in the real-world scenarios and the researchers' approaches to overcome the concerns.
Building our knowledge on top of the initial insights gained from the Gleicher et al.'s survey \cite{gleicher2011visualcomparison}, we suggest relevant empirical evidence we discovered from our survey (e.g., the actual effectiveness of layouts in different study settings) and identify novel facets that the original study did not discover (e.g., the familiarity of layouts to people).
Moreover, we show the consistent and contradictory study results of comparative layouts in terms of different study factors by agglomerating pairwise relations of layouts from \numstudypapers study papers (\autoref{figure:studyresults}).
Going a step further, we suggest \numguidelines actionable guidelines for comparative layouts and promising research directions based on our literature review as well as the results of the \numstudypapers papers.
Finally, we provide a web-based visualization gallery to support designers in exploring the complex design space of comparative layouts using a flexible visualization grammar.
\section{Gleicher et al.'s Comparative layout}
\begin{figure}[!t]
  \centering
  \includegraphics[width=\columnwidth]{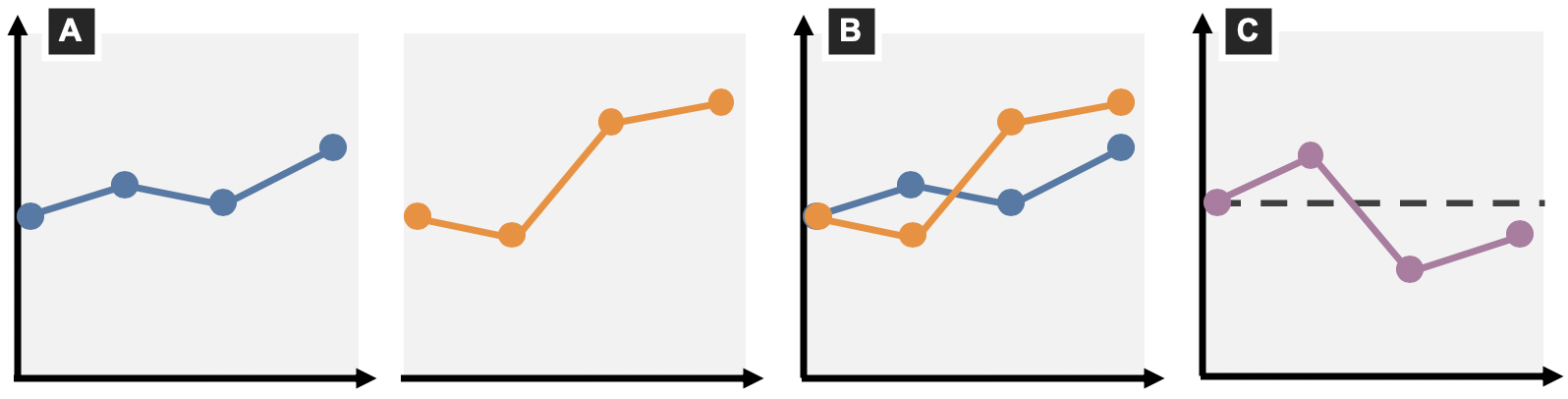}
  \caption{Three primitive information visualization (InfoVis) layouts for visual comparison tasks \cite{gleicher2011visualcomparison}: (A) placing visualizations next to each other (\textit{juxtaposition}), (B) placing visualizations on top of the other (\textit{superposition}), and (C) directly showing the relationship of interests such as subtraction values (\textit{explicit-encoding}).}
  \label{figure:primitive_layouts}
  \vspace{-4mm}
\end{figure}

Throughout the paper, we will use the terms from Gleicher et al. \cite{gleicher2011visualcomparison} to refer to comparative layouts: \textit{juxtaposition}, \textit{superposition}, and \textit{explicit-encoding}.
The three designs describe the arrangement of two or more visualizations to support comparison tasks.
First, juxtaposition refers to placing visualizations next to each other (\autoref{figure:primitive_layouts}A).
It is sometimes called \textit{spatial juxtaposition} to distinguish it from \textit{temporal juxtaposition}, which temporally separates visualizations, for example, switching from one to another or using animated transition \cite{heer2007animated}.
Superposition refers to placing visualizations on top of the other, such as overlaying one line chart on another (\autoref{figure:primitive_layouts}B).
Finally, explicit-encoding focuses on revealing the predefined relationship between visualizations.
For example, if the difference between two trends is of interest, one can explicitly draw the difference on a line chart (\autoref{figure:primitive_layouts}C).
Note that the explicit-encoding layout is not limited to creating a new visualization with aggregated values but also includes visual elements overlaid on the original visualization (e.g., lines connecting the corresponding points in two scatterplots \cite{li2018embeddingvis}).
Designers can also combine multiple layouts (i.e., \textit{hybrid layout}), such as overlapping two node-link diagrams (superposition) with the common edges and nodes highlighted using a different color (explicit-encoding) \cite{naragino2017weightedfreetrees}.
\change{Although animated transition and explicit-encoding are not technically `layouts' (i.e., visualization arrangement), we call them comparative layouts for the consistency with Gleicher et al.'s work.}
\section{Literature Review}
We reviewed \numpapers research papers that employed the comparative layouts to expand our understanding about the layouts.

\subsection{Method}
We reviewed two types of research papers in our survey: (1) \textit{General} papers that employed and/or discussed comparative layouts but have not conducted any quantitative user studies and (2) \textit{Study} papers that conducted quantitative studies to compare the performance of the layouts.

\textbf{General papers.} We firstly looked into all the \numtotalpapers publications that cited the work of Gleicher et al. \cite{gleicher2011visualcomparison} until March 11, 2020 using Google Scholar.
We then excluded irrelevant papers using the following criteria: 
(1) papers which do not explicitly use comparative layouts and do not present any discussions about them (e.g., some papers mentioned comparative layouts only to provide high-level contexts of comparative visualization in introduction), (2) duplicate publications (e.g., thesis papers), and (3) papers written in languages other than English.
Lastly, we excluded (4) papers which mainly focusing on scientific visualization (e.g., 3D blood flow simulation \cite{van2014bloodflow}) to stick to the original focus of the comparative layouts \cite{gleicher2011visualcomparison}, that is information visualization (InfoVis).
\change{However, we did not exclude scientific visualization papers if they suggested any InfoVis techniques with comparative layouts (e.g., juxtaposed line charts~\cite{weissenbock2018dynamic}).}
After the filtering process, we obtained a set of \numgeneralpapers \textit{General} papers.

We surveyed the following factors from the selected papers, which were the main factors discussed in previous papers \cite{gleicher2011visualcomparison, gleicher2017considerations}:

\begin{itemize}
    \item The type of visualizations placed using the layouts
    \item The number of visualizations to compare at once \cite{gleicher2017considerations}
    \item How each of the comparative layouts \cite{gleicher2011visualcomparison} is used
    \item How researchers describe the advantages and concerns of using each layout
    \item Researchers' approaches to overcome the concerns
\end{itemize}

\begin{figure*}
  \centering
  \includegraphics[width=\textwidth]{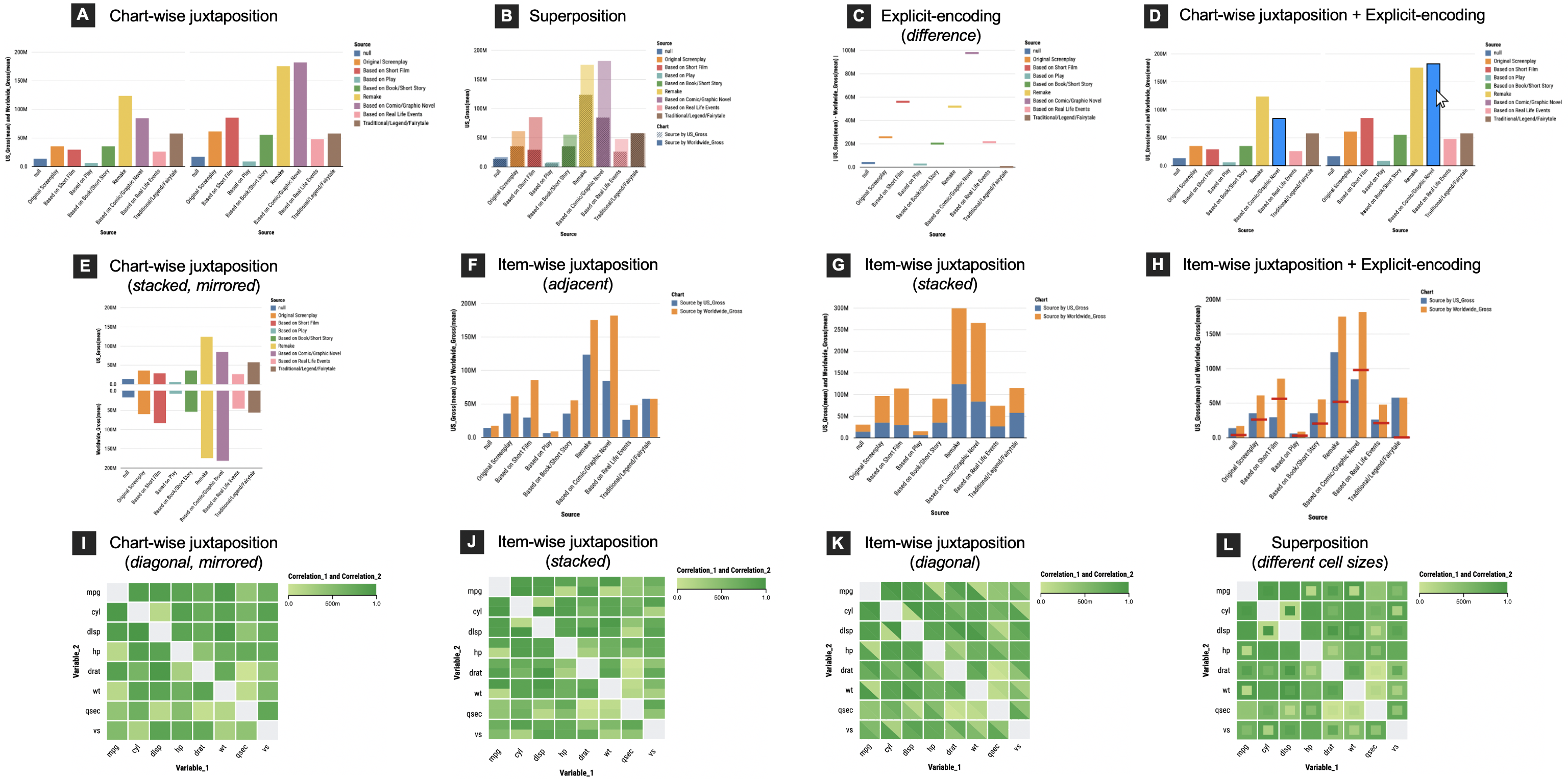}
  \caption{Examples of comparative layouts observed in our literature survey: (A-C) the three primitive comparative layouts (i.e., juxtaposition, superposition, and explicit-encoding); (D) visual linking between adjacently placed bar charts; (E) symmetrically stacked bar charts; (F-G) a grouped bar chart and a stacked bar chart using item-wise juxtaposition; (H) explicit-encoding overlay for showing subtraction values on top of a grouped bar chart; (I-K) variants of chart-wise and item-wise juxtaposition for heatmaps; (L) superposed heatmaps using different cell sizes. \change{Note that the design space of comparative layouts in our classification is not limited to the 12 examples shown in this figure because multiple layouts can be used at once in different combinations, and they can be used in any visualization types.}}~\label{figure:layouts}
  \vspace{-7mm}
\end{figure*}

To avoid ambiguity in collecting the usage of the comparative layouts, we mainly based our data collection on the authors' justifications described in the papers.
Even though visualizations are placed adjacently as in many general visualization systems, we have not regarded this as using a comparative layout (i.e., juxtaposition) unless the authors explicitly stated the proper rationale because it is unclear whether the layout is used for visual comparison or not.
We have not also considered the cases where the different visualization types are placed using the comparative layouts because comparison tasks are most likely to be taken with same visualizations.
One typical example in our review is the difference (explicit-encoding) overlaid on top of a grouped bar chart (juxtaposition) \cite{srinivasan2018whatsthedifference} (\autoref{figure:layouts}H).
In this case, consistent to the authors' explanation, we did not consider it as using an additional superposition layout between the juxtaposed bar chart (\autoref{figure:layouts}F) and the explicit-encoding chart (\autoref{figure:layouts}C) because these two charts are not arranged for comparing the two.

\textbf{Study papers.} To collect the empirical evidence on the performance of comparative layouts, we additionally surveyed the papers that (1) presented quantitative user studies, (2) directly compared the usefulness of the layouts (i.e., comparative layouts being an independent variable), and (3) included visual comparison tasks (e.g., \cite{gleicher2017considerations}).

To complement the relatively small number of papers with quantitative user studies found from the original target papers (i.e., 401 publications that cited Gleicher et al.'s work \cite{gleicher2011visualcomparison}), we additionally looked into two more sets of publications after reviewing the original set.
First, we looked into the \numpapersgleicher papers that Gleicher et al. \cite{gleicher2011visualcomparison} originally reviewed.
Second, we reviewed publications that are published at the following two venues for the last ten years: ACM CHI Conference on Human Factors in Computing Systems (\textit{CHI}) and IEEE Conference on Information Visualization (\textit{InfoVis}).
We chose these two venues since we found the majority (87.5\%) of papers with quantitative user studies were from these venues.
Through this process, we were able to find \numstudypapers \textit{Study} papers.

From the study papers, we collected study conditions (e.g., comparative layouts used as independent variables, tasks of studies, and the number of participants) and results, as well as the main factors we collected from the \textit{General} papers (e.g., advantages and concerns of using each layout).
\def\chalscale{\textit{Chal\textsubscript{scalability}}}

\section{Comparative Layouts in The Wild}
Overall, we found \numdesigns \change{visualization examples from \numpapers papers that used comparative layouts} (about \avedesigns examples per paper).
The most widely used layout is juxtaposition (106), while superposition (39) and explicit-encoding (38) are frequently used as well.
We also found 55 \change{examples} that used multiple layouts at once (i.e., hybrid layout).
The most widely used visualization types include bar charts (64), heatmaps (41), node-link diagrams (40), line charts (39), scatterplots (15), and map visualizations (13).
More than half of the examples (57.4\%) used comparative layouts for comparing a pair of visualizations (i.e., 1:1 comparison).
The \numpapers papers have been published at \numvenues venues;
the majority of papers were from IEEE Transactions on Visualization and Computer Graphics \textit{(TVCG)} (42), ACM CHI Conference on Human Factors in Computing Systems \textit{(CHI)} (11), and Computer Graphics Forum \textit{(CGF)} (10).

\subsection{Classifying Comparison Tasks in User Studies}
\change{Visual comparison tasks are finding and understanding relationship between target visual elements~\cite{gleicher2011visualcomparison,gleicher2017considerations}, such as their similarity or difference, and they usually involve first identifying the target elements in visualizations and then retrieving the relationship between them~\cite{andrienko2006exploratory}}.
For a more comprehensive examination of the \numstudypapers papers with quantitative user studies, we classified comparison tasks (total \numstudytasks tasks) by their scope---\textit{\textbf{global}} (6 papers) and \textit{\textbf{local}} (12 papers)---because we find that these two types of tasks are quite distinguishable in terms of how people perform visual comparison \cite{jardine2019perceptual}.
\textit{Global tasks} refer to comparing the overall characteristics of individual visualizations, such as comparing the correlation of each bar chart.
In contrast, \textit{local tasks} refer to the comparison between visual items, such as comparing the length of bars in two bar charts.
The main characteristic of local tasks compared to global ones is that people must link the corresponding visual elements between visualizations before actually comparing them (e.g., finding bars of the same category in two distant bar charts) unless a system explicitly highlights them.
On the other hand, global tasks require global perspectives that people seem to use more diverse perceptual heuristics in taking the comparison tasks \cite{jardine2019perceptual}.
We used this task categorization to explore the consistent and contradictory study results of comparative layouts (\autoref{figure:studyresults}).

\begin{figure}[!b]
  \centering
  \vspace{-4mm}
  \includegraphics[width=1\columnwidth]{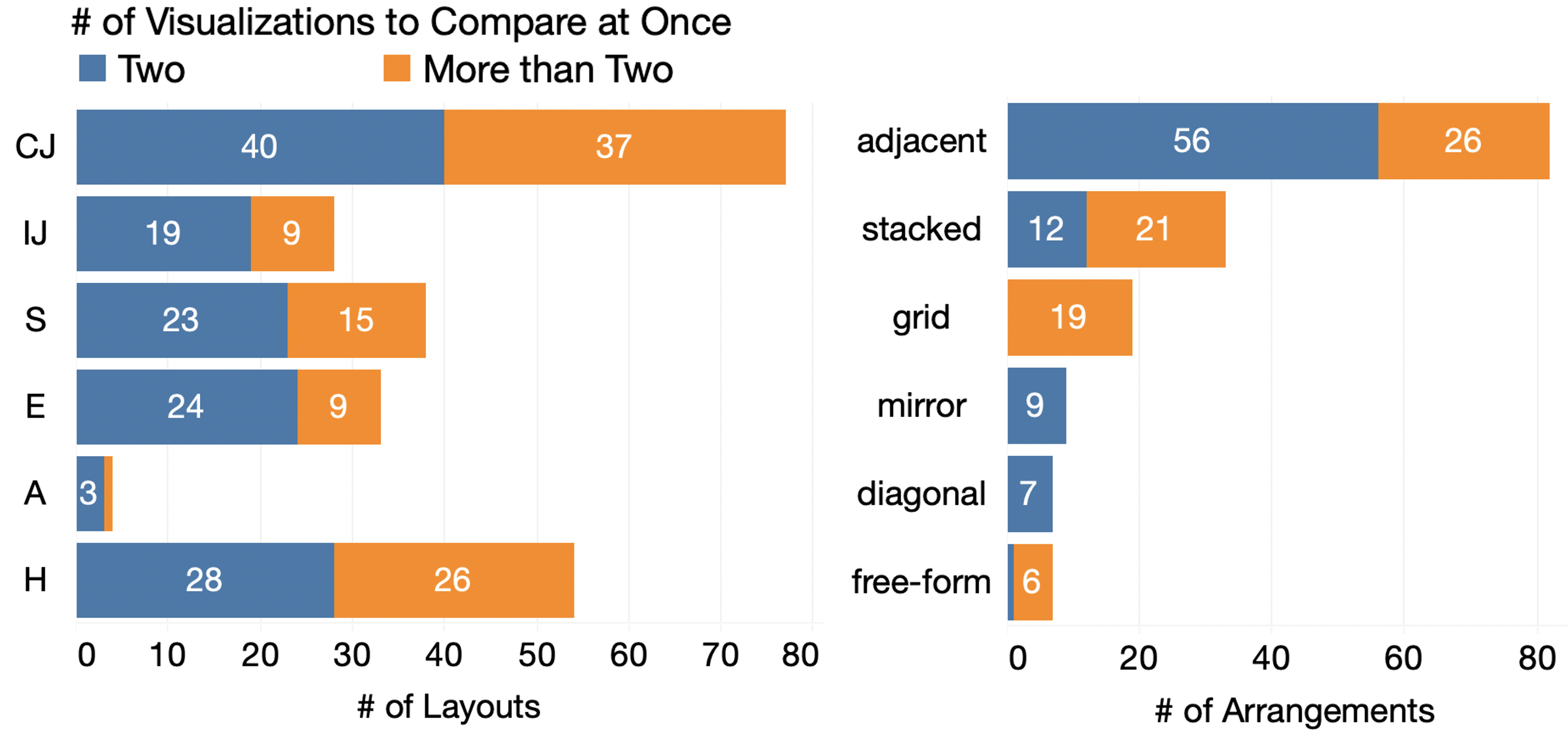}
  \caption{A summary of comparative layouts found from \numpapers papers. The layouts were classified into five exclusive categories: chart-wise juxtaposition (CJ), item-wise juxtaposition (IJ), superposition (S), explicit-encoding (E), and animated transition (A) in addition to an extra hybrid layout (H). The number of layouts are counted exclusively (e.g., superposition used in a hybrid layout is counted only for the (H) hybrid layout). The bar chart on the left shows the number of layouts by the number of visualizations placed together for visual comparison. The chart on the right shows the frequency of arrangements used for juxtaposition.}~\label{figure:num_layouts}
  \vspace{-4mm}
\end{figure}

\subsection{Same Layout Is Called Differently}

We found that the same arrangement of visualizations are often called differently.
One common case is to call a chart with juxtaposed visual marks (e.g., a grouped bar chart; \autoref{figure:layouts}F) either juxtaposition or superposition.
For example, Srinivasan et al. \cite{srinivasan2018whatsthedifference} called a grouped bar chart the juxtaposition layout in that the chart places bars side-by-side.
In contrast, Ondov et al. \cite{ondov2018face} treated the same chart as a superposition layout, considering the chart as multiple bar charts overlaid with different offsets.
Similar problems occur in the case of matrix visualizations  \cite{alper2013weighted,zhao2015matrixwave}.
Temporal juxtaposition, animated transition between multiple charts, is sometimes considered as superposition in that it shows multiple visualizations in the same space \cite{sambasivan2013visualizing,kim20173d4ddata}.
% in the same space while each occupying the same display area (in different times).
Superposition and explicit-encoding are also ambiguous for specific visualization designs.
For example, in the case where two node-link diagrams are shown in a single view with common edges and nodes highlighted, one can consider it as either a single union node-link diagram with explicit-encoding \cite{sambasivan2013visualizing} or superposition of two node-link diagrams with explicit-encoding \cite{naragino2017weightedfreetrees}.

\subsection{Lucid Classification of Comparative Layouts}
To more systematically organize the insights gained in the literature review and provide implications for comparative layouts in a more precise manner without confusion, we found it is necessary to alleviate the ambiguous boundaries between the layouts.
We propose to classify the three comparative layouts into five exclusive categories: (1) chart-wise juxtaposition, (2) item-wise juxtaposition, (3) animated transition, (4) superposition, and (5) explicit-encoding and (6) an extra hybrid layout. \autoref{figure:layouts} shows the examples of each category observed in our target papers, and \autoref{figure:num_layouts} summarizes the overall distribution of each category.

To reflect the diverse variants of juxtaposition layouts, we suggest two subcategories for juxtaposition with six different ways of arrangements.
We classified original juxtaposition into \textit{\textbf{chart-wise}} and \textit{\textbf{item-wise juxtaposition}}, distinguishing the type of targets that are arranged using juxtaposition (i.e., chart or visual elements).
For example, placing two bar charts side-by-side (i.e., concatenating two bar charts) is chart-wise juxtaposition (\autoref{figure:layouts}A), while arranging bars next to each other (i.e., grouped bar charts) is item-wise juxtaposition (\autoref{figure:layouts}F).
Distinguishing item-wise juxtaposition from chart-wise juxtaposition can be especially useful to alleviate possible confusion when discussing their contrasting effectiveness.
For example, we found many papers consider ``juxtaposition'' as the least effective layout for finding small differences, which seemed to refer to chart-wise juxtaposition.
However, item-wise juxtaposition is found to be much more effective than chart-wise juxtaposition according to user study results \cite{ondov2018face,jardine2019perceptual}.
In chart-wise and item-wise juxtaposition, we discovered six different ways of arranging visualizations or visual elements---\textit{adjacent}, \textit{stacked}, \textit{grid}, \textit{mirrored}, \textit{diagonal}, and \textit{free-form} (Fig. \ref{figure:layouts})---where three of the terms are brought from the recent study \cite{ondov2018face}.
Adjacent and stacked arrangements refer to placing charts or visual elements in horizontal and vertical axes respectively, constructing either a grouped or a stacked bar chart in the item-wise version (\autoref{figure:layouts}F and G).
In our survey, several matrix visualizations used diagonal arrangements for these layouts (\autoref{figure:layouts}I and K).
The free-form arrangements are supported when people can interactively rearrange the visualizations without any restrictions.
The mirrored arrangement is placing visualizations symmetrically (\autoref{figure:layouts}E), which can be used with another arrangement where the adjacent arrangement is most frequently used with the mirrored layout.

\textit{Superposition} refers to the designs that combine multiple visualizations into one visualization with a unified coordinate system.
In contrast to chart-wise or item-wise juxtaposition, visual elements can overlap in superposition (e.g., nodes and links can overlap if two node-link diagrams are superposed \cite{alper2013weighted}).
While juxtaposition and superposition refer to static designs, the \textit{animated transition} category refers to the designs that use the temporal transition from one chart to another to highlight the difference between multiple charts. 
The transition usually takes place on the same visualization space, showing a single chart at a time which distinguishes animated transition from juxtaposition or superposition.
\textit{Explicit-encoding} refers to the use of extra visual elements that help comparison.
For example, one can draw lines between two scatterplots to connect the corresponding points \cite{kehrer2013model}) or highlight common edges or nodes between two network diagrams with a different color \cite{naragino2017weightedfreetrees}.
Explicit-encoding can be used without juxtaposition or superposition; for example, if the difference between two bar charts is of interest, one can draw a separate bar chart that only shows the difference without the original bars (\autoref{figure:layouts}C).

In practice, two layouts from different categories can be used together, which refers to \textit{hybrid layout}.
For example, to help people more easily find the related bars in juxtaposed bar charts, systems can visually link them using a difference color (explicit-encoding) upon user interaction (\autoref{figure:layouts}D).
A separate visualization that is constructed using explicit-encoding can be also overlaid on top of juxtaposed bar charts (\autoref{figure:layouts}H) to support accessing both the difference and the original information.
Highlighting common or unique visual elements in superposed node-link diagrams also belongs to this layout.

\begin{figure}[!t]
  \centering
  \includegraphics[width=1\columnwidth]{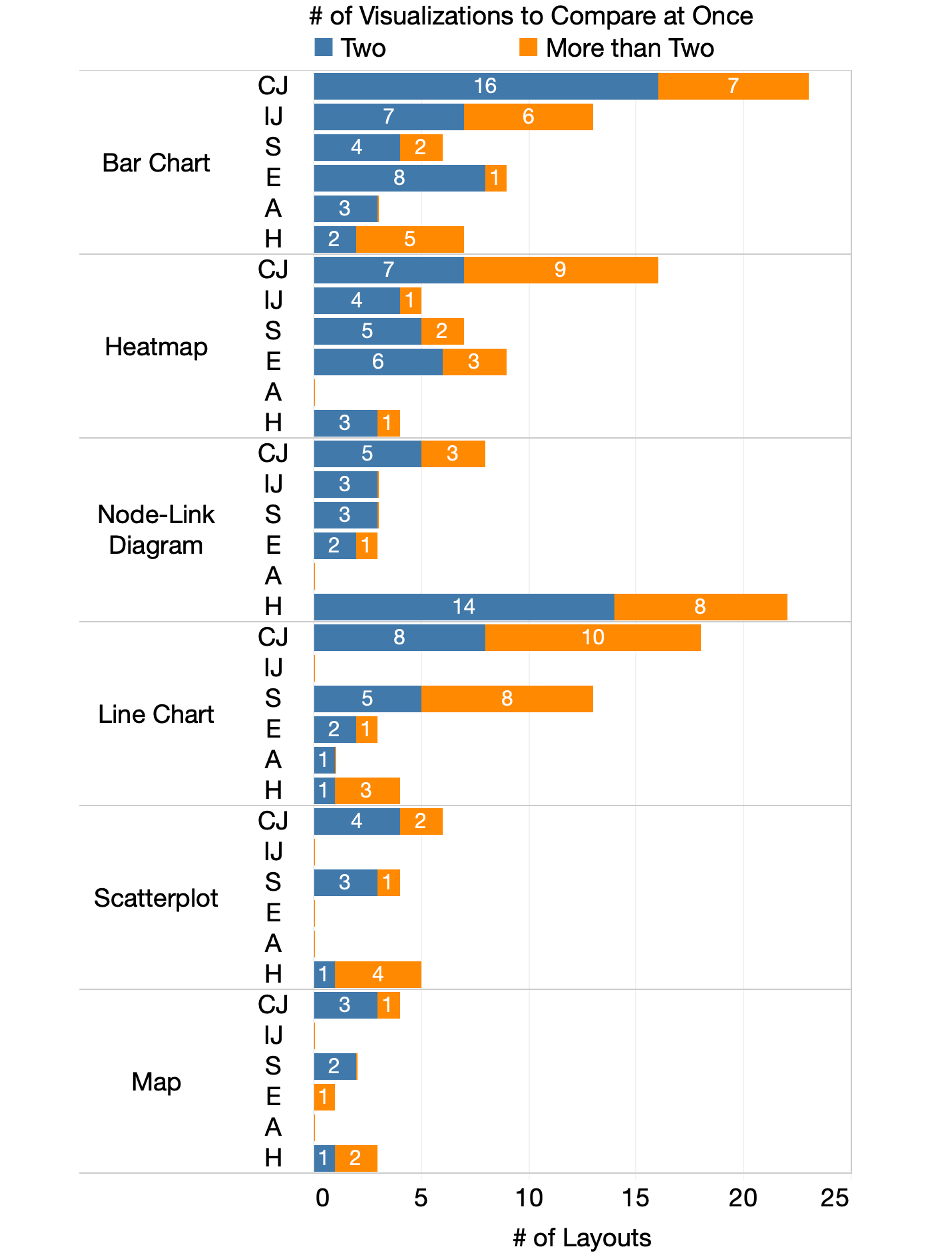}
  \caption{The frequency of comparative layouts grouped by the five most frequent visualization types and the number of visualizations placed together for comparison. 
  %Top five visualization types were selected for this figure based on the frequency from our survey results.
  The abbreviations for comparative layouts are identical to that in \autoref{figure:num_layouts}}
  \label{figure:layout-count}
  \vspace{-5mm}
\end{figure}

\subsection{Uses of Layouts by Visualization Types}
The distribution of comparative layouts depending on visualization types and the number of visualizations placed together is shown in \autoref{figure:layout-count}.
For almost all visualization types, chart-wise juxtaposition is the most frequent.
This may be because of its strong advantages: It is easy to implement \cite{gleicher2011visualcomparison} and supports straight-forward comparison \cite{sambasivan2013visualizing} for small number of visualizations and for comparing non-complex visual stimuli \cite{kehrer2013model}.
An exception to this is when designing node-link diagrams.
Since this type of visualizations is generally more complex compared to bar and line charts, researchers prominently used hybrid layouts to alleviate visual clutters and show differences more clearly \cite{sambasivan2013visualizing,zaman2017mace,rufiange2013diffani}.
Unlike chart-wise juxtaposition, the use of item-wise juxtaposition seems to be restricted by visualization types.
In item-wise juxtaposition, to be able to place corresponding visual elements next to each other, at least one of the $x$ and $y$ axes should encode categorical values.
This seems to be the reason why item-wise juxtaposition is actively used for bar charts and heatmaps while never used in scatterplots and line charts.
Superposition is adopted in the highest proportion compared to other layouts when designing line charts, placing multiple lines in a single view with different colors.
Overall, chart-wise juxtaposition and hybrid layouts are used in higher proportion for comparing more than two visualizations (\autoref{figure:num_layouts}), compared to that of placing only two visualizations, which seems to be a way to overcome the limited scalability that other layouts have.

\begin{table*}[t]
\caption{Various aspects of comparative layouts discussed by researchers in \numpapers papers: advantages and concerns of using the layouts and approaches to overcome the concerns. These aspects (66 categories in total) are categorized into three sub-categories in comparison with the discussions in Gleicher et al.'s work \cite{gleicher2011visualcomparison}: (N) newly found from our survey (35 categories), (E) not new but confirmed by empirical evidence** (15 categories), and (G) neither new nor confirmed* (16 categories).}
\centering
\resizebox{\textwidth}{!}{%
\begin{tabular}{lr l lr l lr l lr}
\multicolumn{5}{l}{\large \textbf{Juxtaposition}} & \textbf{} & \large \textbf{Superposition} &  &  &  & 
\\
\textit{Advantages} & \textit{\#}  & \textit{} & \textit{Concerns} & \textit{\#} & \textit{} & \textit{Advantages} & \textit{\#} & \textit{} & \textit{Concerns} & \textit{\#} \\ \cline{1-2} \cline{4-5} \cline{7-8} \cline{10-11} 
no visual interference* & 8  & & limited scalability** & 12 &  & effective comparison & 4 &  & visual interference* & 17 \\
supporting other tasks & 3  & & difficulty in comparison** & 7 &  & suitable for subtle difference** & 4 &  & limited scalability* & 3 \\
simple implementation* & 2  & & cognitive burden* & 6 &  & easy comparison & 3 &  & visual separation* & 1 \\
straight-forward comparison & 2  & & ineffective comparison** & 5 &  & effective interpretation & 2 &  & lacking intuitiveness & 1 \\
suitable for large difference & 2 & & unsuitable for subtle difference & 4 &  & minimize eye movement* & 2 &  & unfamiliar & 1 \\
easy comparison & 2 &  & difficulty in linking items** & 4 &  & suitable for large difference & 2 &  & precluding other tasks & 1 \\
familiar & 1 &    & managing consistency & 4 &  & less cognitive burden* & 2 &  &  & \\ 
high preference & 1 &  & unsuitable for complex stimuli & 2 &  & suitable for spatial data* & 1 &  &  \textit{Approaches} & \textit{\#}  \\\cline{10-11}
convenient comparison & 1 &  & long eye movement* & 2 &  & high preference & 1 &  &  using hybrid layout** & 2 \\
 & &   & unsuitable for large difference & 1 &  &  &  &  &  aggregating visual elements* & 1  \\
 & &   &  &  &  &  &  &  & managing opacity* & 1  \\
\textit{} & \textit{} & \textit{} & \textit{Approaches} & \textit{\#} & \textit{} & \textit{} & \textit{} & \textit{}  & filter & 1 \\ \cline{4-5}
&  &  & using hybrid layout** & 8 &  & &  &  &  & \\
&  &  & managing consistency & 6 &  & &  &  &  &  \\
&  &  &  shortening distance** & 4 &  & \large \textbf{Animated Transition} &  &   & &  \\
&  &  &  filter & 2 &  & \textit{Advantages} & \textit{\#} & \textit{} & \textit{Concerns} & \textit{\#} \\ \cline{7-8} \cline{10-11} 
&  & &  change arrangement* & 1 &  & suitable for temporal data* &  2 & & no concurrent comparison* & 2  \\ 
%  &  & & &  &  &  &  &  &  &  &  \\
\large \textbf{Explicit-Encoding} & \textbf{} & \textbf{} & \textbf{} & \textbf{} & \textbf{} & effective comparison** &  2 & & cognitive burden* & 2  \\
\textit{Advantages} &   \textit{\#} & \textit{} & \textit{Concerns} & \textit{\#} & \textit{} & structural change &   1& & unsuitable for large difference & 2\\ \cline{1-2} \cline{4-5} 
suitable for subtle difference & 4 &  & information loss** & 4  &  &   showing constancy & 1 & & ineffective comparison** & 2\\
effective comparison** & 4 &  & precluding other tasks** & 3  &  &  showing causality &   1 & & requiring constant attention & 1\\
reasonable scalability & 3 &  & unfamiliar & 1  &  &  showing narratives & 1 & & difficulty in comparison** & 1\\
high preference & 1 &  &  &  &  & supporting other tasks & 1 &  &  & \\ 
& &   &  \textit{Approaches} & \textit{\#}   &  &  &  &  &  \textit{Approaches} & \textit{\#}\\
\cline{4-5} \cline{10-11}
 &   &  &  using hybrid layout** & 1  &  & & & & staging change & 1\\ 
\end{tabular}%
}
\label{table:justifications}
\vspace{-6mm}
\end{table*}

\subsection{Advantages and Concerns of Using Each Layout}

In this section, we reflect on the advantages and concerns of using each layout suggested in the papers to develop our understanding of the comparative layouts in the real-world scenarios (\autoref{table:justifications}).

\subsubsection{Chart-wise Juxtaposition}
The advantages of chart-wise juxtaposition mainly stem from its characteristic that it does not significantly change the original visualization \cite{maries2013grace,lobo2015interactivemap,liu2015effects,correll2018vsup}, which is sometimes the main reason for choosing chart-wise juxtaposition over other layouts \cite{maries2013grace}.
Another related advantage is its ability to support separate analyses of individual visualizations \cite{sambasivan2013visualizing,correll2018vsup,paton2015visualization}, which is an important factor for professional analysts in network analysis \cite{sambasivan2013visualizing}.
Researchers also advocate its applicability to any visualizations \cite{andrienko2018state} or its simplicity in implementation: ``[Juxtaposition is] simple, even trivial'' \cite{beck2017taxonomy}.
When two visualizations are juxtaposed and mirrored, it is known that the human perception system effectively recognizes the symmetry between two visual representations \cite{tyler2003human} which facilitates comparison between the two.
A recent work \cite{ondov2018face} provided practical evidence that juxtaposing two charts in a mirror manner was more efficient than using animated transition or item-wise juxtaposition for comparing the correlation of individual bar charts.

On the other hand, six papers have commonly claimed that the key concern of chart-wise juxtaposition is its limited scalability \cite{gleicher2011visualcomparison, zhao2017mobile, srinivasan2018whatsthedifference, von2018insights, schmidt2013vaico,lobo2015interactivemap}.
For example, it is challenging to juxtapose a large number of visualizations simultaneously in the limited screen space; as an extreme case, it is sometimes impossible to place even two visualizations at the same time in a mobile environment \cite{zhao2017mobile}.
Another concern regarding chart-wise juxtaposition lies in its effectiveness in comparison.
Tominski et al. \cite{tominski2012natural} described this problem as ``eyes have to move from one part to the other part,'' which consequently leads people to rely on the mental image of the first part to compare it with the other part.
In this sense, chart-wise juxtaposition has been criticized for such cognitive cost \cite{tominski2012natural,niederer2017taco,naragino2017weightedfreetrees,liu2015effects} and considered as the least effective layout for comparison tasks compared with other layouts \cite{andrienko2018state,alper2013weighted,cruz2018interactive}.

Specifically, researchers claimed that the subtle difference between juxtaposed visualizations is especially difficult to recognize \cite{ondov2018face,spechtenhauser2018diffpin, wang2018towardseasy ,cruz2018interactive}: ``Spot the difference games, in which observers try to detect small changes ..., illustrate the difficulty of [comparing between two regions]'' \cite{ondov2018face}.
Comparing complex visualizations (e.g., two node-link diagrams) is also claimed to be inefficient \cite{zhao2015matrixwave,kehrer2013model} since people have to temporally remember a complicated representation.
Another concern on chart-wise juxtaposition is that it is difficult to couple the corresponding visual elements from two distant visualizations \cite{correll2018vsup,heimerl2018multiclassscatter,tarner429exploring,lobo2018animation}.
For example, Correll et al. \cite{correll2018vsup} found that people often make mistakes when identifying relevant cells in two heatmaps with chart-wise juxtaposition.
Emphasizing this issue, Lobo et al. \cite{lobo2018animation} claimed that chart-wise juxtaposition can be effective ``only if objects can easily be matched.''
Many researchers also added that, to be effective, designers should carefully optimize the consistency between visualizations \cite{kehrer2013model,bernard2017approaches,dasgupta2015bridging,kim20173d4ddata}, such as using the same range for the axes in chart-wise juxtaposition or placing relevant visual elements in the same logical position in juxtaposed node-link diagrams.

% ``The result shows that com- paring subtle color differences for estimating ratio is difficult.'' [Treemap study]

% \subsubsection{Item-wise Juxtaposition}
% - familiarity \cite{srinivasan2018whatsthedifference}
% % Contrary to widely held belief
% - efficiency \cite{ondov2018face}
% more effective than chart-wise for delta task.

% comparing: The key idea is to dynamically relocate the objects to be compared to form an in-situ juxtaposition. Reducing the distance between the objects makes the visual comparison cognitively easier [PW06].

% - (IJ) With this arrangement, height changes can be identified more effectively, since the respective cuboids are depicted directly next to each other. \cite{limberger201925dtreemaps}

% Another concern for explicit-encoding is its unfamiliarity: ``[V]ariants of familiar chart types are preferred to introducing novel visual encodings'' \cite{srinivasan2018whatsthedifference}.
% For example, if a single bar chart is used to represent values for one series, a variation of a bar chart such as a concatenated, grouped bar chart, or derived difference bar chart would be preferred for comparing values across multiple series, too

\subsubsection{Superposition}
Superposition has been advocated for supporting comparison tasks \cite{niederer2017taco,naragino2017weightedfreetrees, keck2018visual,alper2013weighted,sambasivan2013visualizing}, allowing a ``quick and easy'' comparison \cite{crissaff2017aries}.
Subtle difference, which is challenging to recognize in chart-wise juxtaposition, can be visually salient in superposition \cite{caruso2017creating,cruz2018interactive,heimerl2018multiclassscatter} because target visual elements are arranged closely.
Wang et al. \cite{wang2018towardseasy} argued that superposition is ``especially useful when the spatial location is a key component of the comparison,'' such as in geographical visualizations.
The key concern on superposition is visual interference (i.e., visual elements being overlapped challenge people in interpreting visualizations) which can lead to a scalability issue \cite{kim20173d4ddata,liu2015effects,naragino2017weightedfreetrees,tominski2012natural,vogogias2018bayespiles,von2018insights, zhao2017mobile}.
For example, Viola et al. \cite{viola2017pondering} mentioned the complexity of this concern: 
``[T]he display of several data attributes quickly leads to visual clutter. There is thus no general methodology on how to design effective integrated multi-attribute visualizations.''
% the same reference space implies that data attributes fight for the visual estate at the same locations. Moreover, 
In this context, Caruso et al. \cite{caruso2017creating} asserted that superposition can be useful only when target visualizations are similar enough.
A qualitative study by Tominski et al. \cite{tominski2012natural} showed that it is hard to compare two superposed heatmaps because of the blended color of each cell.
% Another relevant concern is the separability between visual elements from multiple visualizations since people cannot notice which element is from which visualization if the elements are too similar without any distinct visual features (e.g., two scatterplots overlapped using the same shape and color for individual points).
% In addition, Zhao et al. \cite{zhao2017mobile} mentioned that superposition is less intuitive compared with chart-wise juxtaposition because chart-wise juxtaposition does not change individual visualizations.

\subsubsection{Explicit-Encoding}
The main advantage of explicit-encoding is that it allows direct access to the predefined relationship \cite{niederer2017taco,zhao2015matrixwave, tominski2016comparing}: ``[T]he viewer does not need to make a mental comparison or find the difference, as it has already been calculated'' \cite{niederer2017taco}.
For this reason, explicit-encoding can be used for designs where visualizing subtle difference is of importance \cite{krueger2016traveldiff}.
Its second advantage is the scalability in terms of the number of target visualizations since it usually focuses only on showing the predefined relationships without showing the original visualization.
For example, in a mobile environment, explicit-encoding can be more effective than juxtaposition or superposition \cite{zhao2017mobile} since the screen space is limited.
Based on user studies, researchers also found explicit-encoding is useful when overlaid with other layouts (i.e., hybrid layouts).
The hybrid layouts allowed a faster and more accurate comparison between node-link diagrams \cite{naragino2017weightedfreetrees} and were more preferred by people \cite{srinivasan2018whatsthedifference} compared with using a single layout.

However, it can be ineffective if people can only see a specific relationship without the original information: ``Ideally, we would like to see the entire dataset without missing any detail, but explicit-encoding concedes this design goal ... in favor of others'' \cite{kim20173d4ddata}.
This seems a considerable drawback as data analysts described in a research paper \cite{dasgupta2015bridging} did not like such information abstraction: ``[D]ue to information loss, scientists were not comfortable with the idea of smoothing by computation of average.''
% Similarly, Naragino et al. \cite{naragino2017weightedfreetrees}  emphasized that ``preparing explicit-encoding for each task is not a very efficient method in a series of tasks.''

A relevant problem of explicit-encoding is called \textit{decontextualization}, which involves losing contexts of data in visual representations: ``The user sees the result of a comparison but cannot interpret it without additional visualization of the original data. This increases the complexity of the visualization'' \cite{von2018insights}.
% Another concern for explicit-encoding is its unfamiliarity: ``[V]ariants of familiar chart types are preferred to introducing novel visual encodings'' \cite{srinivasan2018whatsthedifference}.
% For example, if a single bar chart is used to represent values for one series, a variation of a bar chart such as a concatenated, grouped bar chart, or derived difference bar chart would be preferred for comparing values across multiple series, too
Another concern for explicit-encoding is its unfamiliarity.
A study with treemap visualizations \cite{limberger201925dtreemaps} showed that people occasionally misinterpreted a novel textual representation that encodes the direction of value changes.
Similarly, participants from another study had difficulties in interpreting explicitly encoded differences (\autoref{figure:layouts}F), and they rated explicit-encoding least effective compared with item-wise juxtaposition or hybrid designs \cite{srinivasan2018whatsthedifference}.

% explicit encoding does not guarantee more accurate difference measurement, but it appears to reduce the time to complete the task.
\subsubsection{Animated Transition}
% The animated transition is regarded as an effective method for showing causality and narratives in a series of visualizations \cite{lobo2018animation}.
Animated transition is especially useful for recognizing a small local difference between two visualizations, as it outperformed item-wise and chart-wise juxtaposition in finding the maximum difference between a pair of bar charts or donut charts \cite{ondov2018face}.
Because animated transition shows visualizations separately in time, it allows people to take independent analyses on each visualization \cite{sambasivan2013visualizing}.
% Animated transition is known to be effective when showing time series data \cite{lobo2018animation,cruz2018interactive}.
However, the drawback of this layout is that people cannot see target visualizations at once \cite{kim20173d4ddata,srinivasan2018whatsthedifference}, which is known to be less effective than comparing concurrently visible representations \cite{munzner2015visualization} especially when the number of target visualizations increases.
Moreover, animation requires constant attention and interaction (e.g., switching between views repeatedly) \cite{ondov2018face, zhao2015matrixwave,alper2013weighted}, which ``may increase the time requirement'' \cite{alper2013weighted}.
The performance of animated transition on comparison tasks is controversial; while animated transition showed outstanding performance in a study \cite{ondov2018face} with an local task, it resulted in inaccurate comparison even with confusion with node-link diagrams \cite{sambasivan2013visualizing}.
Similarly, experts who used animated scatterplots to see multiple t-SNE results mentioned that watching animated transition was cognitively challenging: ``[T]racking the nodes in an animated manner requires a mental map comparison, which is demanding'' \cite{li2018embeddingvis}.
\subsection{Approaches to Overcome the Concerns}
In this section, we discuss researchers' previous attempts to overcome the concerns of each layout to develop deeper insights of comparative layouts with diverse design options.

\subsubsection{Chart-wise Juxtaposition}
We found four main approaches for chart-wise juxtaposition to overcome its limited scalability and ineffectiveness in comparison tasks.
% For item-wise juxtaposition, we were not able to 
% Because item-wise juxtaposition is used as alternative designs for other layouts for complementing 

\textit{Using Hybrid Layout.}
Explicit-encoding is frequently used to complement chart-wise juxtaposition \cite{correll2018vsup,heimerl2018multiclassscatter,kim20173d4ddata,vogogias2018bayespiles,krueger2016traveldiff}.
We identified two major purposes of this approach: (1) assisting to couple the corresponding visual elements and (2) improving the effectiveness in the recognition of difference.
For example, egoComp \cite{liu2017egocomp} used lines connecting visual elements in multiple visualizations ``to decrease the user's memory cost.''
Heimerl et al. \cite{heimerl2018multiclassscatter} suggested explicitly showing bin boundaries in multi-class scatterplots to ``[h]elp with mapping bins across different plots.''
% Correll et al. \cite{correll2018vsup} claimed the necessity of highlighting the corresponding cells between two heatmaps with chart-wise juxtaposition.
To address the difficulty in comparing a large number of heatmap visualizations in chart-wise adjacent arrangements, BayesPiles \cite{vogogias2018bayespiles} allowed people to select a reference heatmap to temporally color-encode differences (i.e., subtraction values) in the rest of the matrices.
% Interaction Techniques for Visual Exploration Using Embedded Word-Scale Visualizations.
Results from user studies \cite{srinivasan2018whatsthedifference,naragino2017weightedfreetrees} support the effectiveness of hybrid layouts as using explicit-encoding overlays with chart-wise and item-wise juxtaposition in bar charts and node-link diagrams showed better performance compared with solely relying on the juxtaposition layouts (\autoref{figure:studyresults}A and E).

\textit{Shortening Distance.}
Juxtaposing visualizations or visual elements as close as possible is one of the simplest but effective methods.
A body of studies showed empirical evidence that comparison is easier when visual representations are closer together \cite{larsen1998effects,talbot2014four,plumlee2006zooming}.
We identified four studies that explicitly mentioned using similar approaches \cite{birkeland2014perceptually,srinivasan2018whatsthedifference,tominski2016comparing,tominski2012natural}:
``When the two stimuli are far away from each other, the subject has to frequently move the eyes to switch the focus. Therefore, ... we have placed the stimuli as close to each other as possible'' \cite{birkeland2014perceptually}.
With user interaction, Tominski et al. \cite{tominski2012natural} allowed people to crop and bring the rectangular part of a visualization close to the area to which they want to compare it with.
We also found two studies that used item-wise juxtaposition for this purpose;
for example, Srinivasan et al. \cite{srinivasan2018whatsthedifference} ``opted to use a grouped bar chart instead of a concatenated bar chart (bar charts with chart-wise juxtaposition) since comparisons are likely to be more accurate with no distracting bars in between corresponding values.''
In a geographical visualization, CompaRing \cite{tominski2016comparing} brings a few regions of comparison candidates near a reference region upon user selection.
Study results support the effectiveness of item-wise juxtaposition \cite{naragino2017weightedfreetrees,sambasivan2013visualizing,ondov2018face} in enhancing comparison performance in terms of time and accuracy, especially in local tasks.

\textit{Maintaining Consistency.} 
Gleicher et al. \cite{gleicher2011visualcomparison} mentioned the importance of maintaining the consistency of visual properties in chart-wise juxtaposition to minimize cognitive burden.
This is relevant to consistency management in multiple coordinated views \cite{qu2017keeping}, such as determining whether to use shared or independent data domains and ranges on the screen for individual visual channels (e.g., color, size, and the $x$ and $y$ axes).
Likewise, Kim et al. \cite{kim20173d4ddata} mentioned, ``[Keeping visualizations consistent] seems to be particularly useful for juxtaposition because they provide a common context to link the data instances.''
Examples include arranging categories in the same order between heatmaps \cite{zhang2014visual} or using a constant height for all visualizations \cite{herr2018visual}.
% For comparing multiple videos, Tharatipyakul and Lee \cite{tharatipyakul2018towards} devised a method called content synchronization, or temporal alignment, that shows similar objects in individual videos at a time point.
We also found that almost all studies that employed chart-wise juxtaposition used this approach by using a constant color scheme \cite{tominski2016comparing}, size \cite{tominski2012natural}, and coordinate systems \cite{zeng2017cnncomparator}.

\textit{Filter.} 
The number of items or visualizations being compared simultaneously is known to determine the difficulty in comparison tasks \cite{gleicher2017considerations}.
% In this sense, applying filters on data or visualizations can be effective: ``Looking at the related work in juxtaposition as a whole, it is evident that visualization designers often employ user interaction ... in order to reduce an originally large number of data instances to a manageable subset ...'' \cite{kim20173d4ddata}.
For example, CompaRing \cite{tominski2016comparing} automatically selects a few number of comparison targets to reduce the complexity, and Zaman et al. \cite{zaman2017mace} proposed ``subtractive encoding'' which removes common nodes and edges from network visualizations to highlight the differences.

\subsubsection{Superposition}
We discuss two approaches to alleviate the main drawback of superposition, visual interference.
% as well as support visual separability in superposition.

\textit{Using Clutter Reduction Methods.}
To manage the visual interference, clutter reduction methods can be employed, which can be categorized into Ellis et al.'s taxonomy of clutter reduction techniques \cite{ellis2007taxonomy}.
For example, Dasgupta et al. \cite{dasgupta2015bridging} aggregated multiple lines as a band to prevent them from being a ``spaghetti plot.''
Many studies controlled the transparency \cite{tominski2012natural} or size \cite{alper2013weighted} of visual elements, while filtering visual elements \cite{zaman2017mace} is also a popular method.
Other methods include jittering or adding offsets along axes in line charts \cite{dasgupta2017empirical} and node-link diagrams \cite{naragino2017weightedfreetrees}.
% To use superposition in multi-class scatterplots, Heimerl et al. \cite{heimerl2018multiclassscatter} used blending and weaving techniques.

\textit{Using Hybrid Layout.}
Although not commonly suggested, complementing superposition using explicit-encoding seems promising to overcome the visual interference and further enhance its performance in comparison tasks.
For example, inspired by natural behaviors with printed papers, one study \cite{tominski2012natural} allowed people to peek at the summary of occluded regions through a folding interaction and found that this kind of explicit-encoding on demand complements the weakness of superposition.
Similarly, VAICo \cite{schmidt2013vaico} used explicit-encoding in superposed images to summarize and show the clusters of inconstant regions with user interactions.
Another result shows that highlighting common or unique nodes and edges in superposed node-link diagrams outperformed a single layout with few exceptions \cite{naragino2017weightedfreetrees} and were preferred by professionals \cite{sambasivan2013visualizing}.
% \textit{Managing Inconsistency.}
% To make multiple visualizations visually separable in superposed visualization, most superposition designs attempted to distinguish the visual elements from different visualizations:
% ``[W]e use colour hues ... to reinforce the distinction between perspective views'' \cite{beecham2016faceted}.
% Color channels (i.e., hue, saturation, and luminance) were commonly used to encode the memberships of target visualizations, for example, in line charts \cite{ming2017understandinghidden}, scatterplots \cite{shen2017streetvizor}, and bar charts \cite{malik2016high}, while the use of different textures and shapes was observed in a few studies \cite{kehrer2013model,piringer2012comparative} (\autoref{figure:layouts}B).

\subsubsection{Approaches for Other Layouts}
% The major concern regarding explicit-encoding was the information loss that obstructs accessing the original visualizations, that is, decontextualization.
In explicit-encoding, researchers used hybrid layouts to complement the weaknesses of explicit-encoding (i.e., decontextualization and unfamiliarity).
One study \cite{marti2015hcmapper} discussed this issue and suggested using additional layouts as a remedy: ``To avoid decontextualization using only explicit-encoding ..., we also use juxtaposition.''
A similar approach was evaluated in a study \cite{srinivasan2018whatsthedifference} that using a single explicit-encoding chart showed least preference by the unfamiliarity, but when used with an item-wise juxtaposed visualization, the preference became the best compared to other variants of bar charts.
For animated transition, the use of staged changes between spatial locations is advocated, as the animation often confused people when transition between two visualizations with a large amount of difference took place \cite{sambasivan2013visualizing}.

\begin{figure*}
  \centering
  \includegraphics[width=\textwidth]{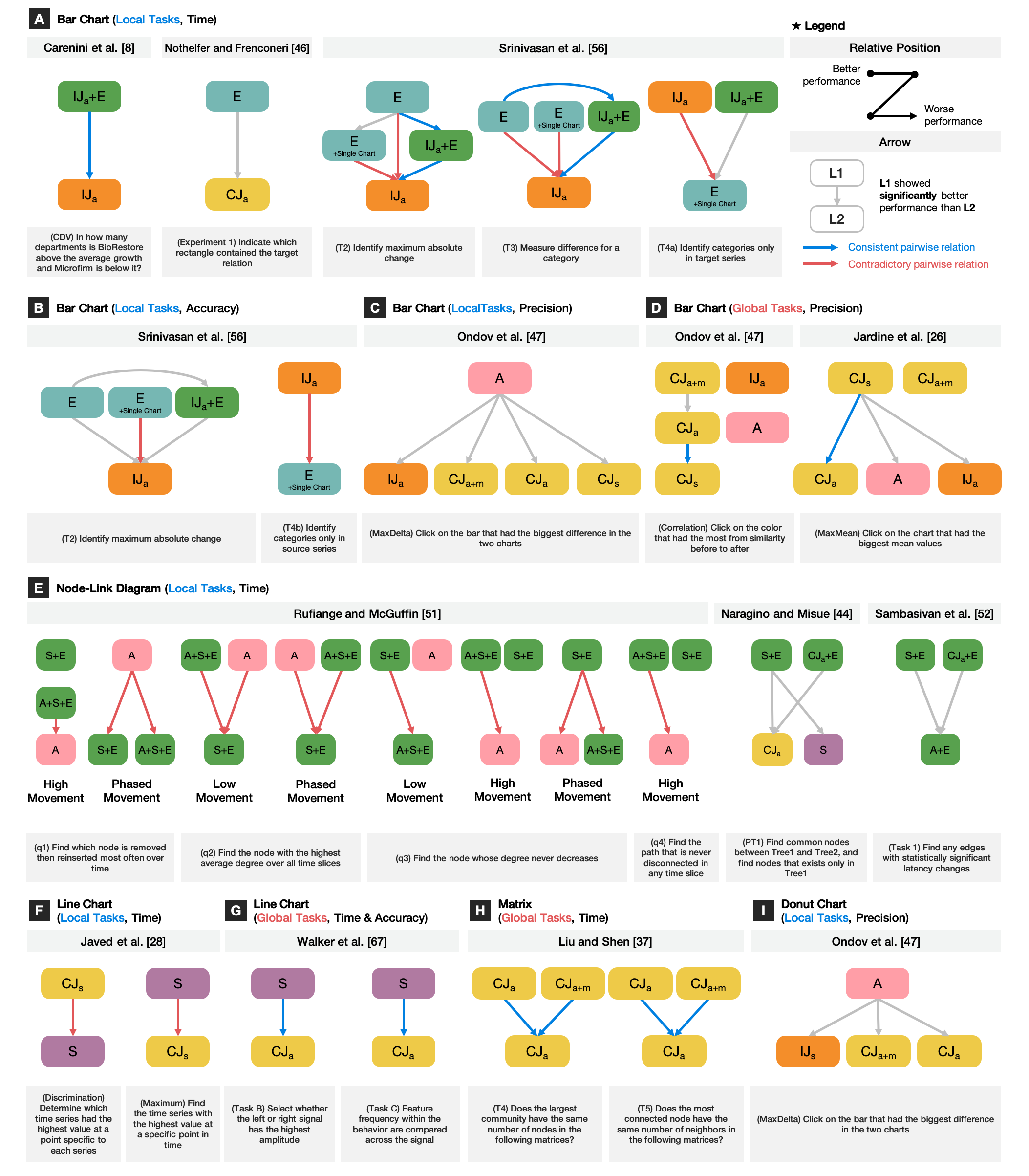}
  \caption{A visual summary of the consistent and contradictory pairwise relations of comparative layouts found in \textit{Study} papers \cite{carenini2014highlighting,nothelfer2019measures,srinivasan2018whatsthedifference,ondov2018face,jardine2019perceptual,rufiange2013diffani,naragino2017weightedfreetrees,sambasivan2013visualizing,javed2010graphical,walker2015timenotes,liu2015effects}. The diagrams (A-I) are arranged in terms of six visualization types, two task types (\textit{local} and \textit{global} tasks), and three dependent variables (\textit{completion time}, \textit{accuracy}, and \textit{precision}) where the \textit{precision} refers to tier values in a staircase method \cite{ondov2018face,jardine2019perceptual}. Arrows indicate that the source is significantly more effective than the destination. The comparative layouts, represented as nodes, are arranged in a way that layouts with better performances (not significantly better without arrows) are placed from top to bottom and left to right. The blue and red edges indicate the consistent and contradictory relations in a certain combination of visualizations, tasks, and dependent variables, respectively, and gray edges indicate that the relation between the two layouts is discovered only once in that combination. The abbreviations of comparative layouts used in this figure are identical to that in \autoref{figure:num_layouts} except that subscripts in juxtaposition layouts indicate the different arrangements: \underline{a}djacent, \underline{s}tacked, \underline{d}iagonal, and \underline{m}irrored. The full summarization of study results can be found at \url{https://sehilyi.github.io/comparative-layout-explorer/}.}~\label{figure:studyresults}
\end{figure*}

\subsection{Comparative Layout Explorer}
To better help designers systematically explore the design options of comparative layouts, we provide a web-based visualization gallery.
It shows diverse designs that are observed in the literature review, such as the visualizations in \autoref{figure:layouts}, and enables people to flexibly change the layout based on the following visualization grammar: 
\newline\newline
{\small\fontfamily{cmtt}\selectfont
\noindent\code{\hspace{0.3cm}Layout :=} Type, Unit, Arrangement, Mirrored

\noindent\code{\hspace{0.3cm}Type :=} Juxtaposition | Superposition | Explicit-Encoding

\noindent\code{\hspace{0.3cm}Unit :=} Chart | Item

\noindent\code{\hspace{0.3cm}Arrangement :=} Adjacent | Stacked | Diagonal | Animated

\noindent\code{\hspace{0.3cm}Mirrored :=} True | False
}
\newline\newline
People can select one of three comparative layouts (i.e., juxtaposition, superposition, and explicit-encoding) and test the diverse ways of arranging visualizations in juxtaposition:
the unit of comparison targets (chart-wise or item-wise), arrangements (adjacent, stacked, diagonal, or animated), and the use of mirrored arrangements.
Since visual consistency and interference are important factors for comparative layouts according to our survey results, we allow users to configure them as well, such as using shared, independent, or distinct color palettes between juxtaposed bar charts or using a different size for cells in superposed heatmaps.
This visualization gallery can be accessed via \url{https://sehilyi.github.io/comparative-layout-explorer/}.

\section{Discussion}
We offer practical implications for comparative layouts by suggesting their trade-offs, actionable guidelines, and promising directions for future research.

\subsection{Trade-offs between Comparative Layouts}

To assist designers in selecting comparative layouts, we summarize and reshape our findings discussed in the previous sections into trade-offs of using the four most frequently used layouts---chart-wise juxtaposition (CJ), item-wise juxtaposition (IJ), superposition (S), and explicit-encoding (E)---in terms of four main themes: scalability, effectiveness in recognizing a predefined relationship, familiarity, and supporting visualization tasks other than comparison.
We present a general consensus made by researchers in the parentheses next to the name of each theme, where ``\textit{T (L1 $>$ L2)}'' represents that the \textit{L1} layout is commonly said to be better than \textit{L2} in terms of the \textit{T} theme, and $\approx$ represents that their effectiveness depends on situations.

\textit{Scalability (E $>$ CJ $\approx$ IJ $\approx$ S)}.
Explicit-encoding is commonly regarded as the most scalable layout for the increasing number of target visualizations because it focuses only on a specific relationship.
This seems a strong advantage for explicit-encoding since the other three layouts are commonly complained of their limited scalability.
For this reason, explicit-encoding was favored by researchers when dealing with small screen space or a large number of visualizations \cite{zhao2017mobile,fofonov2019projected,tominski2016comparing}.
However, the scalability of the rest seems to depend on other factors such as screen space availability and visual representation complexity, leading to the consideration between space efficiency and visual interference.

\textit{Effectiveness in Recognizing a Relationship (E $>$ S $\approx$ IJ $\approx$ CJ)}.
Researchers commonly claimed that recognizing a specific relationship is most effective with explicit-encoding because it directly calculates and represents the relationship \cite{srinivasan2018whatsthedifference,nothelfer2019measures}.
Between the rest, though the general consensus is that shorter distance between comparison targets is more effective, we found chart-wise juxtaposition is sometimes more effective in global tasks compared with item-wise juxtaposition \cite{ondov2018face,jardine2019perceptual}.
Therefore, their effectiveness may split depending on what relationship people are dealing with.

% In addition, researchers generally acknowledged that superposition is more effective than juxtaposition because visual elements are kept closer in superposition.
% For this reason, researchers commonly claimed that small or complex difference between visualizations is easier to be noticed in superposition \cite{keck2018visual,cruz2018interactive,caruso2017creating}.
% However, we found that this general consensus is not entirely correct, and the task performance can be mixed depending on what relationship people are dealing with.
% For example, Ondov et al. \cite{ondov2018face} showed that two bar charts using chart-wise juxtaposition with a mirrored design outperformed item-wise juxtaposition when the task was comparing overall characteristics of two visualizations (i.e., correlation).

\textit{Familiarity (CJ $>$ IJ $\approx$ S $>$ E)}.
Although it may depend on the visualization types used, chart-wise juxtaposition seems to provide the most familiar visualization to people because it does not require any significant modification to individual visualizations.
Between item-wise juxtaposition and superposition, neither seems to entirely outperform the other as we find both the familiar and unfamiliar examples for each layout: Grouped bar charts and multi-class scatterplots can be considered as familiar visualizations of using item-wise juxtaposition and superposition respectively, while variants of heatmaps \cite{zhao2015matrixwave} and node-link diagrams \cite{alper2013weighted} are the unfamiliar ones.
Explicit-encoding is likely to provide the least familiar outcomes because it frequently employs novel visual representations with data aggregation, which is known to be unfamiliar to InfoVis novices \cite{grammel2010information}.

\textit{Supporting Other Types of Tasks (CJ $>$ IJ $\approx$ S $>$ E)}.
Because visual analytics involves performing a series of multiple tasks, the importance of supporting other types of tasks, as well as comparison tasks, is emphasized by many researchers \cite{tominski2016comparing,naragino2017weightedfreetrees,srinivasan2018whatsthedifference}.
The consensus in this respect is that explicit-encoding is least effective since it generally eliminates the original visualizations \cite{srinivasan2018whatsthedifference,dasgupta2015bridging}.
On the other hand, chart-wise juxtaposition is commonly claimed to support general tasks the best by separately showing individual visualizations.
We do not yet have any clear understanding between item-wise juxtaposition and superposition in supporting independent analysis, which hence requires additional studies.

\subsection{Guidelines for Comparative Layouts}
% By combining the literature review and the results of \numstudypapers papers with quantitative user studies, we suggest \numguidelines actionable design implications for comparative layouts.

\subsubsection{Avoid Chart-wise Juxtaposition for Local Comparison}
When looking into the studies with \textit{local tasks} \cite{naragino2017weightedfreetrees,ondov2018face,sambasivan2013visualizing,srinivasan2018whatsthedifference}, %(i.e., comparing visual elements in visualizations)
chart-wise juxtaposition has \textit{\textbf{almost never}} outperformed any other layouts in terms of accuracy, completion time, and precision \cite{ondov2018face,jardine2019perceptual} (see the yellow \textit{CJ} nodes in \autoref{figure:studyresults}).
Considering the diverse factors used in the studies (e.g., visualization types, stimuli complexity, data size, and amount of difference), these consistent results give a very strong implication that if detecting local differences is the main task, designers must use other layouts, such as item-wise juxtaposition or superposition.
This implication is aligned with other existing studies \cite{larsen1998effects,talbot2014four,plumlee2006zooming}, but we confirm it in the context of comparative layouts by categorizing tasks into global and local comparison.

\subsubsection{If Chart-wise Juxtaposition is Inevitable, Provide Landmarks}
In the study results for local comparison tasks \cite{lobo2015interactivemap,naragino2017weightedfreetrees,ondov2018face,sambasivan2013visualizing,srinivasan2018whatsthedifference}, we found a few exceptions where chart-wise juxtaposition showed comparable results to item-wise juxtaposition or superposition.
The first case was when target visual elements are highlighted (i.e., using explicit-encoding additionally) so that people did not have to manually link them \cite{naragino2017weightedfreetrees} (e.g., the last diagram in \autoref{figure:studyresults}E).
The second case was when dealing with geographical visualizations of showing dense regions so that some kinds of landmarks already existed, for example, buildings and roads, which people can use when identifying the corresponding visual elements \cite{lobo2015interactivemap}.
Therefore, it is desirable to provide landmarks, such as grids or reference lines, or further using explicit-encoding to enhance the performance; however, please note that providing landmarks in chart-wise juxtaposition did not make dramatic performance improvements to outperform item-wise juxtaposition and superposition.

\subsubsection{Make Sure to Align Comparison Targets in Superposition}
Aligning comparison targets looks to be the fundamental design choice to take advantage of the effectiveness that superposition potentially has.
This is beyond simply using consistent value ranges for $x$ and $y$ axes in multiple visualizations \cite{qu2017keeping} (e.g., using an identical date period for two trends in a superposed line chart) but is to more actively align `visual marks' that need to be compared.
% Popular examples are superposed node-link diagrams discovered in our survey \cite{}; Even though 
The study results with line charts well illustrate this tendency (\autoref{figure:studyresults}F-G):
Although superposed line charts overall showed superior performance than chart-wise versions, comparing values between different time points for different trends made the superposed line charts significantly ineffective compared to the chart-wise one (see the ``Discrimination'' task).
If the comparison targets can be predefined, they can be aligned in the visualization by default, or interactive designs that allow people to use various alignments on demand can be adopted \cite{goffin2020interaction}.

\subsubsection{Avoid Blending Colors for Superposition}
According to an observation study~\cite{tominski2012natural}, using superposition in heatmaps was not effective because people had difficulties in distinguishing the blended color of cells.
Consistent with the observation, we found only a few examples of using superposition for heatmaps without any techniques for reducing the visual interference.
To prevent such a blending problem, visualization designers can use one of six alternative methods that we discovered in our review.
For comparing a pair of heatmaps, first, designers can simply use glyph visualizations \cite{tominski2012natural}, such as encoding the radius of circles rather than their color.
Second, if two quantitative values are orthogonal (e.g., value and uncertainty), designers can consider using different color channels, such as hue and saturation, following a successful design in uncertainty visualizations \cite{correll2018vsup}.
Third, superposing heatmaps with different cell sizes can be an effective design as several studies showed \cite{alper2013weighted,zhao2015matrixwave} (\autoref{figure:layouts}L).
Fourth, instead of superposition, variants of item-wise juxtaposition can be used with stacked or diagonal arrangements \cite{alper2013weighted,zhao2015matrixwave} (\autoref{figure:layouts}J and \autoref{figure:layouts}K).
Lastly, when the number of visualizations becomes larger, weaving techniques \cite{heimerl2018multiclassscatter,hagh2007weaving,luboschik2010new} or using explicit-encoding to reveal the accurate difference in chart-wise juxtaposed heatmaps \cite{vogogias2018bayespiles} can be used.

\subsubsection{Avoid Solely Using Explicit-encoding}
Explicit-encoding seems to be the most delicate layout, which has strong advantages and weaknesses at the same time.
Despite its effectiveness in recognizing the predefined relationship as advocated by many researchers, others suggested its strong drawbacks.
One example is its unfamiliarity, which can affect InfoVis novices in learning and interpreting visualizations \cite{grammel2010information}.
Explicit-encoding has commonly received low preference to InfoVis novices \cite{sambasivan2013visualizing,srinivasan2018whatsthedifference} and often showed poor performance by the unfamiliarity \cite{limberger201925dtreemaps}.
Moreover, by the decontextualization, it is often criticized by professionals in the real-world scenarios \cite{dasgupta2015bridging,sambasivan2013visualizing}.
%, which reflects the weight of drawbacks that explicit-encoding has.
Since many researchers suggested strong rationales when they had to use explicit-encoding in their paper (e.g., perceptual advantages \cite{vogogias2018bayespiles} or scalability \cite{zhao2017mobile}), we think it should be used when its advantages are certain and surpass its diverse shortcomings.
One such example would be using explicit-encoding for alleviating perceptual distortions in superposed line charts such as Playfair's chart~\cite{tufte2001visual}.

\subsubsection{If Explicit-encoding is Necessary, Use a Hybrid Layout}
According to our review, a hybrid layout seems to well complement the disadvantages that a single layout has, and it was one of the most frequently used layout in our target papers.
All the user studies (four out of \numstudypapers) that used hybrid layouts showed superior performance for specific tasks compared with a single layout, such as effectiveness in detecting and measuring local changes \cite{sambasivan2013visualizing,naragino2017weightedfreetrees,schmidt2013vaico}, high preference \cite{srinivasan2018whatsthedifference,sambasivan2013visualizing}, and better scalability \cite{schmidt2013vaico}.
The study results with bar charts are good examples that illustrate the ability of hybrid layouts for complementing other layouts (\autoref{figure:studyresults}A-B): Solely using an explicit-encoding layout (\autoref{figure:layouts}C) was least preferred by people and sometimes showed the worst performance depending on tasks, but placing it on a familiar grouped bar chart (i.e., item-wise juxtaposition; \autoref{figure:layouts}H) made it most preferred by people and was never the worst.
% Moreover, the only user study \cite{srinivasan2018whatsthedifference} that compared a explicit-encoding layout with its hybrid version well explains the ability of hybrid layouts for complementing other layouts: Although solely using explicit-encoding (\autoref{figure:layouts}C) was least preferred by people, placing it on a familiar grouped bar chart (\autoref{figure:layouts}H) made it most preferred and showed the best performance with the comparable results.
We think that to protect comparative visualizations from the strong weaknesses that explicit-encoding have, designers should consider using other layouts together.

\subsubsection{Refrain from Using Animation for Large Difference}
One study showed that animated transition showed the best performance for detecting a small difference in item-wise comparison, outperforming all other layouts (i.e., chart-wise and item-wise juxtaposition) \cite{ondov2018face}.
However, its performance seems very sensitive to tasks, visualization types, and visual complexity (\autoref{figure:studyresults}E).
For example, in global tasks such as identifying the maximum correlation \cite{ondov2018face} and structural changes \cite{naragino2017weightedfreetrees}, the performance became weaker.
Moreover, a large amount of changes between two node-link diagrams \cite{sambasivan2013visualizing} confused people, leading to poor task performance in accuracy.
As a remedy, designers can consider using staged animation \cite{heer2007animated}, which was helpful for large changes.
However, we still identify many unexplored areas for this layout in visual comparison tasks (e.g., task types, visual representations, and data complexity).
As it showed relatively large performance variations across different designs, designers should use animated transition with care and refrain from using it for detecting large differences.

\subsection{Promising Directions for Future Research}

\change{\textit{Researching Human Factors in Chart-wise Comparison.}}
As we lack empirical results for the performance of global comparison tasks (six out of \numstudypapers papers), exploring comparative layouts with diverse global tasks seems a promising direction to expand our understanding of the layouts.
When we looked into the study results with local comparison tasks, we were able to find relatively consistent results among comparative layouts.
However, it seems that for global tasks, the task performance is much more sensitive, even sensitive to different ways of using chart-wise juxtaposition (\underline{a}djacent, \underline{s}tacked, and \underline{m}irrored in \autoref{figure:studyresults}D).
For example, using mirrored and adjacent chart-wise juxtaposition showed the best performance in correlation tasks \cite{ondov2018face}, but for comparing mean of individual visualizations \cite{jardine2019perceptual}, a stacked arrangement was the best.
As recent work suggested \cite{jardine2019perceptual}, different perceptual heuristics of people seem to greatly influence the performance, resulting in varying performance by target relationships or visual representations.

\change{\textit{Investigating the Effectiveness with Varying Difference.}}
In our review, one of the factors that researchers most frequently discussed for their designs was the amount of difference in terms of size or complexity.
For example, chart-wise juxtaposition is generally regarded as least effective for detecting a small difference because of the long distance between visual elements.
However, the performance might depend on what kinds of small difference users are dealing with, either a global or a local difference, as chart-wise juxtaposition performed better than item-wised juxtaposition for a certain task \cite{ondov2018face,jardine2019perceptual}.
As we find none of the user studies in our survey directly confirmed these aspects by varying size or complexity of difference, it looks worth exploring research topic.

\change{\textit{Investigating the Scalability of Comparative Layouts.}}
Most user studies (nine out of \numstudypapers) focused only on one-to-one comparison.
However, in the real world, more than two visualizations are frequently compared together \cite{gleicher2017considerations}.
Juxtaposition and superposition are considered to suffer from limited scalability, compared to explicit-encoding \cite{kim20173d4ddata,tominski2012natural,vogogias2018bayespiles,von2018insights, zhao2017mobile, lobo2015interactivemap}.
% - about complex \cite{sambasivan2013visualizing}
Therefore, although an independent use of explicit-encoding was the least preferred design for comparing only the small number of visualizations \cite{srinivasan2018whatsthedifference}, it might show the opposite results when the number of visualizations increases to some extent.
To develop a better understanding of comparative layouts in the real world, it looks promising to investigate the ability in terms of scalability.

\subsection{Limitations}
The main reason that we focused on the papers that cited Gleicher et al.’s work \cite{gleicher2011visualcomparison} was to make the data collection process reproducible, accurate, and efficient.
By this choice, we could clearly define the scope of target papers with reasonable quantity and coverage (\textit{N}=\numpapers, \textit{Venues}=\numvenues). 
Furthermore, since these papers often explicitly discussed the strengths or weaknesses of the layouts through the terminology that is consistent with the work of Gleicher et al., we could also gather and amalgamate their insights more accurately.
For example, there are a number of visualization interfaces that place visualizations side-by-side, but not all of them may be designed for visual comparison.
However, we believe that it would also be interesting and worth looking into the `outside' of our scope by reviewing a broader set of papers.
For example, animated transition is relatively less discussed in our paper due to the limited number of papers that employed this layout.
Since there might be many research papers that employ animated transition in comparison contexts without citing the Gleicher at al.'s work, it would be worth to further explore this layout to strengthen our understanding of comparative layouts.
\section{Conclusion}
We presented a systematic review of \numpapers research papers, including \numstudypapers papers with quantitative user studies, to better understand the three comparative layouts for visual comparison: juxtaposition, superposition, and explicit-encoding.
Combining and systematizing the insights previously gained \textit{in the wild}, we offered a consistent and reusable framework for using comparative layouts.
We explored the diverse aspects of comparative layouts, including the advantages and concerns of each layout, approaches to overcome the concerns, and trade-offs between them.
We also proposed \numguidelines actionable guidelines and unexplored research area to reveal promising future directions.

\acknowledgments{
This work was supported by the National Research Foundation of Korea (NRF) grant funded by the Korea government (MSIT) (No. NRF-2019R1A2C2089062).
The ICT at Seoul National University provided research facilities for this study.
Jaemin Jo was affiliated with Seoul National University at the time of this research and is currently affiliated with Sungkyunkwan University.}

\bibliographystyle{abbrv-doi}

\bibliography{reference}
\end{document}